\documentclass[aps,prb,reprint,superscriptaddress,showpacs]{revtex4-1}

\usepackage{graphicx}
\usepackage{times}
\usepackage{amsmath}
\usepackage{subfigure}
\usepackage{epstopdf} 

\begin{document}

\bibliographystyle{aip}

\title{A Numerical Model of Crossed Andreev Reflection and Charge Imbalance}

\author{J.L.~Webb}
\email[email:~]{}
\homepage[web:~]{http://www.stoner.leeds.ac.uk}
\affiliation{School of Physics \&\ Astronomy, University of Leeds,
Leeds, LS2 9JT UK}

\author{B.J.~Hickey}
\affiliation{School of Physics \&\ Astronomy, University of Leeds,
Leeds, LS2 9JT UK}

\author{G.~Burnell}
\email[email:~]{G.Burnell@leeds.ac.uk}
\affiliation{School of Physics \&\ Astronomy, University of Leeds,
Leeds, LS2 9JT UK}

\date{\today}% It is always \today, today,

\begin{abstract}
We present a numerical model of local and nonlocal transport properties in a lateral spin valve structure consisting of two magnetic electrodes in contact with a third perpendicular superconducting electrode. By considering the transport paths for a single electron incident at the local F/S interface - in terms of probabilities of crossed or local Andreev reflection, elastic cotunneling or quasiparticle transport - we show that this leads to nonlocal charge imbalance. We compare this model with experimental data from an aluminum-permalloy (Al/Py) lateral spin valve geometry device and demonstrate the effectiveness of this simple approach in replicating experimental behavior. 
\end{abstract}

\maketitle

\section{\label{sec:level1}Introduction}

Crossed Andreev reflection (CAR) is a charge transfer process whereby an electron incident at a normal metal or ferromagnet to superconductor junction may enter the superconductor at energies less than the gap energy $\Delta$ through simultaneous retroreflection of a hole in a spatially separate, nonlocal electrode. Previous work by others has demonstrated this effect experimentally \cite{2004PhRvL..93s7003B}$^{,}$ \cite{2005PhRvL..95b7002R} for a nonlocal or lateral spin valve geometry consisting of two separate normal state electrodes incident with a third superconducting electrode, laterally separated on the scale of the BCS coherence length $\xi_0$. This effect has been observed as a negative nonlocal voltage V$_{nl}$ and negative nonlocal differential resistance dV$_{nl}$/dI, where V$_{nl}$ is measured across the second normal electrode (defined as the detector electrode) separate to that through which current I is applied (the injector) \cite{2007ApPhA..89..603B}. CAR has attracted considerable recent interest due to the potential to create solid state quantum entanglement via the splitting of a Cooper pair via CAR, demonstrated by recent experimental studies \cite{2010NatPh...6..494W}$^{,}$\cite{2009NatPh...5..393C}.

Previous work has shown that other competing processes also contribute to the nonlocal effect - elastic cotunneling (EC) by which an electron may tunnel nonlocally via an intermediate virtual state in the superconductor and nonlocal charge imbalance (CI), an effect produced through the creation of charge nonequilibrium in the superconducting electrode through quasiparticle injection and diffusion \cite{2010PhRvB..81r4524H}. Both of these processes have been suggested to fully or partially cancel the CAR effect, by producing an equal but opposite contribution in nonlocal voltage, or to dominate over CAR in the regime near T$_c$ or near the critical field. The use of ferromagnetic electrodes, magnetized parallel(P) or antiparallel(AP), has been suggested and demonstrated as a potential means to separate the spin-dependent CAR and EC effects \cite{2000ApPhL..76..487D}

A number of theoretical models have been developed to model the lateral spin valve by taking an analytical approach - either solutions of the Bogoliubov-de Gennes equations, following the method of Blonder, Tinkham and Klapwijk(BTK) \cite{1982PhRvB..25.4515B}$^{,}$\cite{PhysRevB.68.174504} for Andreev reflection at a single N/S junction, or by solution of the Usadel equations \cite{2006PhRvB..74u4512B}$^{,}$\cite{2009PhRvL.103f7006G}. Such studies suffer the deficiency of being untested against experimental data or being unable to simultaneously fully replicate all the nonlocal effects - particularly those due to nonlocal charge imbalance and the negative nonlocal resistance associated with CAR. In this paper we present a simplified means to model these effects by considering the possibilities for a single electron incident at the local (injector) N/S interface and demonstrate the effectiveness in such an approach in terms of replicating real experimental data. 

\begin{figure}
\includegraphics[width=1.0\linewidth]{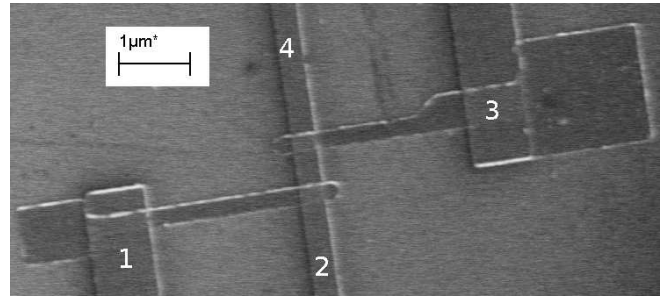}\\
\caption{SEM image of device. Py magnetic electrodes (1) and (3) intersect with an orthogonal Al electrode (2)-(4). Current application and voltage detection is performed as in the text }
\label{fig:SEMimage}
\end{figure}

\section{\label{sec:level1}Experiment} 

In order to provide comparison with the model, devices were fabricated for experimental measurement by electron beam lithography using a Raith 50 system in a standard lateral spin valve configuration. These consisted of two Ni:Fe 80:20 (Py) electrodes with 1x1$\mu$m and 2x2$\mu$m nucleation pads and attached nanowire with either a continuous 300nm in width or tapering from 600nm to 300nm (over 400nm), contacting a perpendicular 300nm width Al electrode, of lateral inter-electrode separation 600-900nm (Fig. \ref{fig:SEMimage}). Patterning was performed using standard PMMA positive resist. Deposition of Py was performed by DC magnetron sputtering at 34W/2.5mTorr Ar at system base pressure 3x10$^{-9}$torr. Overlay of the Al electrode was by EBL overlay patterning followed by Al deposition at 50W/2.5mTorr Ar at base pressure 3.2x10$^{-8}$torr, preceded by an in-situ 40s Ar$^{+}$ mill at 410V acceleration voltage to clean the Py electrode surface of oxide and residual resist. Growth thicknesses of the Py and Al electrodes were 15nm and 30nm respectively for all devices. Measurement was undertaken in an adiabatic demagnetization refrigerator cooled to 600mK with current injection and local voltage detection between 1-2 in Fig. \ref{fig:SEMimage} and nonlocal voltage detection 3-4, using both a DC applied current I$_{dc}<$100$\mu$A and nanovoltmeter and AC current $I_{ac}$=0.25$\mu$A with DC offset $I_{dc}$ in order to measure differential conductance via a standard lock-in amplifier technique. Low AC frequency of 9.99Hz was used to minimize inductively generated nonlocal voltage observed.

\subsection{\label{sec:level2}Local Effect}

Differential conductance through the local (injector) junction was measured using the lock-in method with I$_{ac}$=0.25$\mu$A and DC offset I$_{dc}$ to 50$\mu$A. The resulting data can be seen in Figure \ref{fig:local1} and exhibited a subgap enhancement of conductance consistent with single junction Andreev reflection \cite{jr.:4589}. The model of Blonder, Tinkham and Klapwijk (BTK) and as modified by Strijkers et. al. \cite{1982PhRvB..25.4515B}$^{,}$\cite{2001PhRvB..63j4510S} was fitted to the experimental data by a standard $\chi$-squared minimization process to obtain parameters (see figure caption) including spin polarization P, interface parameter Z and effective temperature T$^{*}$ that were within physical limits. Two methods of fitting were used: with a single gap parameter $\Delta_1$ and two parameters $\Delta_1$ representing a region of suppressed superconductivity and $\Delta_2$, a higher (bulk) value. Although the single-$\Delta$ approach was capable of replication of the peak at zero and of high I$_{dc}$ behavior, only the two-$\Delta$ approach was able to replicate the finite bias minima observed at I$_{dc}$=10-30$\mu$A in the conductance data. This implied the existence of a region of suppression of the superconductivity in the Al - or a proximity effect induced superconducting region in the injector electrode - potentially arising from the effect of the adjacent magnetic electrode. The low value of P compared to the P=0.3-0.4 range expected can be attributed to the non-point contact nature of the junction and that the polarization returned reflects the effective polarization at the junction rather than the bulk value for the Py. 

\subsection{\label{sec:level2}Nonlocal Effect}

\begin{figure}
\centering
\includegraphics[width=1.0\linewidth]{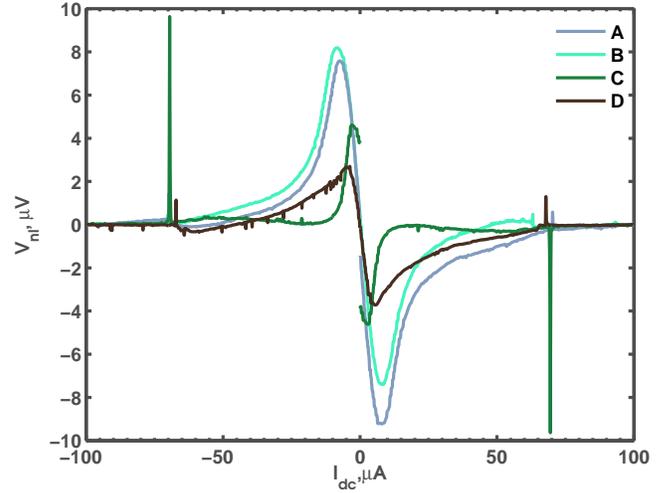}\
\caption[]{Nonlocal DC voltage $V_{nl}$ vs. $I_{dc}$ at 600mK for 4 devices A,B,C,D of inter-electrode separation A=150,B=250,C=600,D=900nm}
\label{fig:DC1}
\end{figure}

\begin{figure}
\centering
\includegraphics[width=1.0\linewidth]{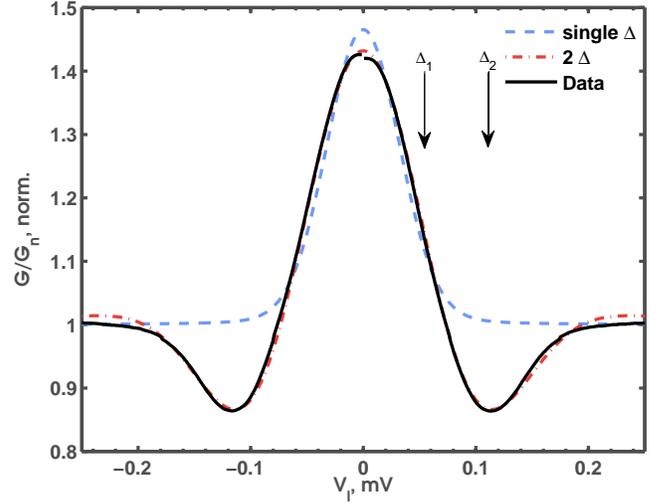}\
\caption[]{Local normalized differential conductance G/G$_n$ for an example injector junction, with BTK model fits using a single-$\Delta$ ($\Delta_1$) and by a two-$\Delta$ ($\Delta_1$ and $\Delta_2$) method. Fitting parameters of P=0.182, $\Delta_1$=0.1491meV, $\Delta_2$=0.2892meV, Z=0.145 and T$^{*}$=1.7K were used, the position of the $\Delta_1$ and $\Delta_2$ indicated on the figure.}
\label{fig:local1}
\end{figure}

\begin{table} 
\centering
\begin{tabular}{ | l | l | l|l|l|l|}
\hline
Device& Separation L,nm & $\Delta_1$(meV) & $\Delta_2$(meV) & Z & P\\
\hline
\hline
Device A & 150nm & 0.120 & 0.254 & 0.051 & 0.135\\
Device B & 250nm & 0.119 & 0.261 & 0.120 & 0.114\\
Device C & 600nm & 0.115 & 0.242 & 0.051 & 0.136\\
Device D & 900nm & 0.114 & 0.240 & 0.024 & 0.139\\
Device E & 600nm & 0.148 & 0.288 & 0.145 & 0.182\\
\hline
\end{tabular}
\caption{Summary of device properties. $\Delta_1$ and $\Delta_2$ values are those of the injector junction, obtained by fitting to the model of BTK. Also included are the spin polarization P, barrier parameter Z and inter-electrode separation L.}
\label{tab:dprop}
\end{table}

\begin{figure}
\centering
\includegraphics[width=1.0\linewidth]{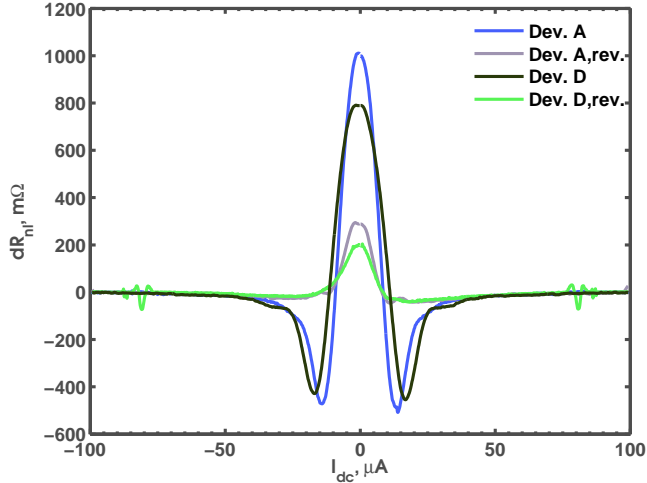}\
\caption[]{Nonlocal resistance $dR_{nl}$ measured via the lock-in technique at T=600mK as a function of $I_{dc}$ at the injector for Device A  of inter-electrode separation 250nm and Device D of inter-electrode separation 900nm. Measurements are shown for both configurations of injector/detector; the reverse (\textit{rev}) configuration refers to current injection through the untapered (300nm width) electrode}
\label{fig:dR1}
\end{figure}

Examples of nonlocal measurement data taken for a range of devices can be seen in Figures \ref{fig:DC1}-\ref{fig:D}. Figure \ref{fig:DC1} shows the nonlocal DC voltage $V_{nl}$ measured for a range of devices A-D of inter-electrode separation L 150-900nm measured as a function of applied DC current through the injector $I_{dc}$. We include a range of properties for these devices in Table \ref{tab:dprop}. A negative nonlocal voltage was observed, with a general trend of reduction in amplitude with increasing inter-electrode separation L characteristic of a CAR effect decaying over $\xi_0$. This effect can also be seen in a plot of nonlocal differential resistance $dR_{nl}$ in Figs \ref{fig:dR1} for two of the devices A, D demonstrating a characteristic shape of finite bias negative minima and zero $I_{dc}$ peak, attributed to EC and CAR-dominated transport regimes \cite{PhysRevLett.95.027002}$^{,}$\cite{2010PhRvB..81b4515B}. The effect was observed to be stable on repetition and highly dependent on the injector/detector selection, reversing the choice producing a lower $dR_{nl}$ effect likely arising from variation in local junction properties of the injector.  The absence of a finite $dR_{nl}$ at high $I_{dc}$ was potentially due to the elimination of nonlocal CI effects at the 600mK measurement temperature that would otherwise contribute a finite nonlocal voltage.

\subsection{\label{sec:level2}Temperature Dependence}

\begin{figure}
\centering
\includegraphics[width=1.0\linewidth]{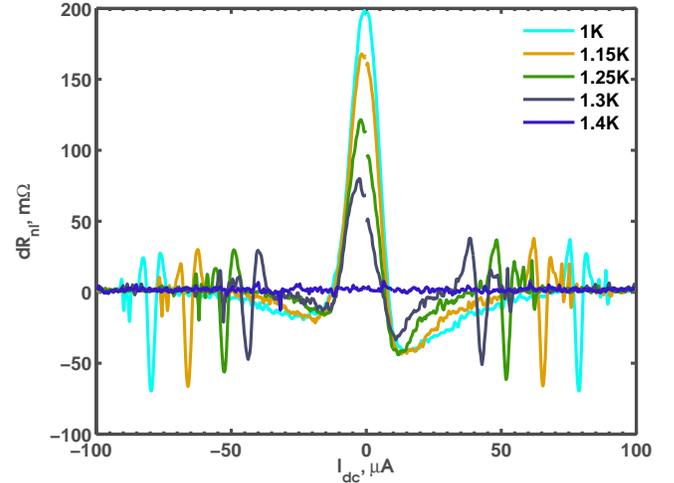}\
\caption[]{Device C nonlocal resistance $dR_{nl}$ as a function of $I_{dc}$ at a range of temperatures. No signal was detected at T$\ge$1.4K, with T$_c$ for the device reached at $\approx$1.35K. The asymmetry in $\pm$I was a product of the measurement method}
\label{fig:tmp1} 
\end{figure}

\begin{figure}
\includegraphics[width=1.0\linewidth]{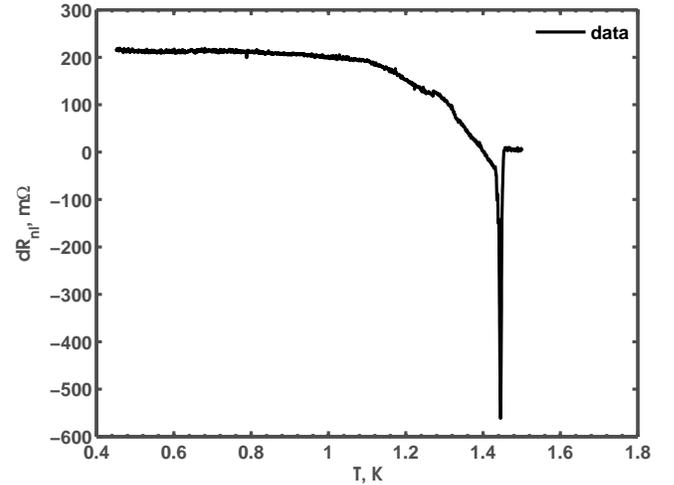}
\caption{Experimental temperature dependence of nonlocal differential resistance $dR_{nl}$ for a separate Al/Py Device E of inter-electrode separation 900nm, with slightly higher T$_{c}\approx$1.4K}
\label{fig:D}
\end{figure}

Figure \ref{fig:tmp1} shows the effect of changing temperature on $dR_{nl}$ vs. I$_{dc}$ for a separate Device E with L=600nm. No effect was observed above T$_c$ (at T$>$1.3K in the figure). A reduction in both peak height and negative minima depth was observed, due to reduction in nonlocal superconducting effects. A shift towards $I_{dc}$=0 of the zero bias minima was observed, likely due to reduction in $\Delta$ with increasing T. At higher $I_{dc}$ sharp inversions in $dR_{nl}$ were observed, the position of which reduced in I$_{dc}$ as temperature increased. We attribute these features to nonlocal charge imbalance, given the temperature and current dependence and similarity in shape to features observed by others- notably in the the work by Cadden-Zimansky and Chandrasekhar\cite{2006PhRvL..97w7003C}. 

Figure \ref{fig:D} shows the variation in zero bias (I$_{dc}$=0) peak with temperature from 0.45-1.5K for a separate device. A finite, temperature independent effect was observed as T$\rightarrow$0 which we attribute to the elimination of nonlocal charge imbalance due to the reduction in the charge imbalance length $\Lambda_Q$ = $\sqrt{D\tau_Q}$, with D the metal diffusion constant and $\tau_Q$ the charge imbalance time given by:

\begin{equation}
 \tau_Q = \frac{4k_BT}{\pi\Delta(T)}\tau_{in}
\end{equation}

\noindent
with $\tau_{in}$ the inelastic scattering time. 

At low temperatures, the CI effect is minimized, implying the existence of other non-CI effects (i.e. - EC or CAR) to create the finite peak. As T$\rightarrow$T$_c$, this finite value is reduced to zero with a large inversion in $dR_{nl}$ close to T$_c$. No effect was observed above T$_c$, as expected if the measurements were attributable to nonequilibrium superconducting processes. We attribute this inversion effect also to nonlocal charge imbalance, arising from the divergent increase in $\Lambda_Q$. Such an inversion effect has been seen in studies of local and nonlocal CI effects\cite{2007NJPh....9..116C}$^{,}$\cite{2007NJPh....9..116C}. Previous works by Beckmann et al. \cite{2004PhRvL..93s7003B} and Kleine et al. \cite{2010Nanot..21A4002K} have however shown a different form for temperature dependence at zero bias, an increase from a low or zero nonlocal resistance at low temperature to a finite positive peak as T$\rightarrow$T$_c$, followed by a sharp drop to zero above T$_c$. We attribute the difference in this work to be due to the presence of an additional finite effect at low temperature, observed as a peak at zero bias and likely arising from EC, and to a differing negative contribution to nonlocal resistance arising from either nonlocal CI or device properties such as the effect of a suppressed T$_c$ at the interface region. Further discussion of this aspect is given in a later section. 

\section{\label{sec:level1}Modeling}

In order to understand the experimental effects seen here and elsewhere in the literature, we consider a numerical model of the processes undertaken by a single electron incident at the injector. We initially model the incident electron as a single charge packet, interacting only with the electric field in the wire and undergoing diffusive motion with a mean free path $\lambda_F$ in the injector (ferromagnetic) material and $\lambda_S$ in the normal (or superconducting) metal. An initial position for the electron is defined at $y$=-100$\lambda_F$ with the inter-metallic interface at $y$=$y_0$=0. We define positions $x_1$ and $x_2$ as the left and right-hand edges of the injector electrode and $y_1$ the width of the orthogonal electrode, such that for y$<$0 the electron is confined within $x_1$ to $x_2$. We also define position $y_1$ as the top edge of the normal metal electrode, such that the electron is confined to the region y$\leq$$y_1$. 

For the electron dynamics, we set an initial electron velocity $\textbf{v}_0$ from a Fermi distribution function at system temperature T at a point $x_0$ where $x_1<x_0<x_2$, $y$=-100$\lambda_F$ and direction is defined by unit velocity vector $\hat{\textbf{v}}$ such that:

\begin{equation}
\textbf{v}_0 = |\textbf{v}_0|\hat{\textbf{v}}= \begin{array}{c}1 \\ \sqrt{v_{x0}^2 +v_{y0}^2}\end{array} \left(\begin{array}{c} v_{x0}\\v_{y0}\end{array}\right)
\end{equation}

The values of position $x_0$ and component of $\hat{\textbf{v}}$ are taken from a pseudo-random number generator. The system is allowed to iterate with time step $\Delta t$, with time $t$=n$_t\Delta t$ for n$_t$ timesteps. The electron acceleration $\textbf{a}$ due to the electric field in the wire, the velocity $\textbf{v}$ and new position are calculated at each time step, giving a time-dependent position $\textbf s$:

\begin{equation}
\left(\begin{array}{c} s_{x}\\s_{y}\end{array}\right) = \left(\begin{array}{c} x_{0}\\-100\lambda_F\end{array}\right) +  \left(\begin{array}{c} v_{x0}\\v_{y0}\end{array}\right)t + \frac{1}{2} \left(\begin{array}{c} a_{x}\\a_{y}\end{array}\right)t^{2}
\end{equation}

We set the simulated time step $\Delta t$ such that the distance traveled by the electron on each step $|\textbf{s}| = \sqrt{|\Delta s_x|^{2}+|\Delta s_y|^{2}}<<\lambda_F$, in practice 1-2nm so as to reduce computation time. We define a scattering time $\tau_F$ where $\lambda_F$=$|\textbf{s}|\tau_F$; when $t$=n$\tau_F$, with n an integer, the electron velocity is reset from the distribution function and new pseudo-random values set for the direction vector $\hat{\textbf{v}}$. 

\begin{figure}
\includegraphics[width=1.0\linewidth]{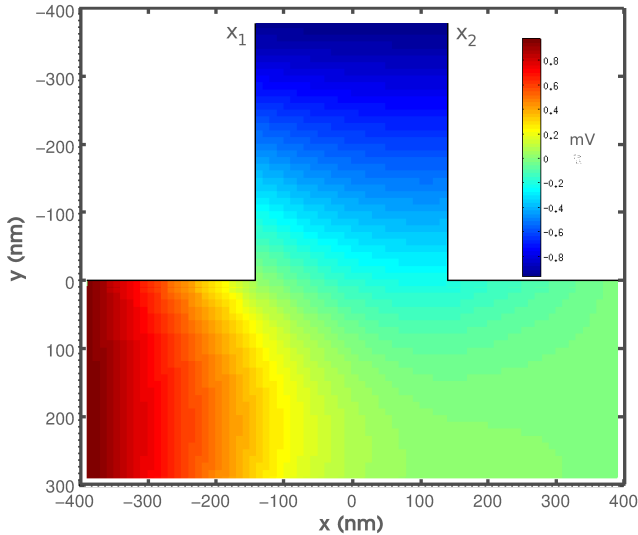}
\caption{Electric potential $\phi$ numerically calculated in the device at V=0.1mV as a function of position $x$ and $y$, from which a spatially varying electric field $\textbf{E}(x,y)$ = ($\partial\phi$/$\partial x$, $\partial\phi$/$\partial y$) could be calculated }
\label{fig:dpath}
\end{figure}

We calculate the electric field $\textbf{E}(x,y)$ in the structure from the derivative of the numerical solution of the Laplace equation for the electric potential $\phi$, including the relative resistivities of the metal electrodes and invoking the following boundary conditions, where V is the applied bias to the injector: \\ 

\noindent

\begin{equation}
\phi= \left\{\begin{array}{c l}
  -V & \mbox{$x$$_1\leq$$x$$\leq$$x_2$, $y$=-100$\lambda_F$} \\
  +V & \mbox{$x$=-$x_{lim}$, $y\ge$0} \\
   0 & \mbox{$x<x_1$ or $x>x_2$, $y=0$ or $y=y_1$}\\
\end{array}
\right.
\end{equation}

\begin{equation}
\frac{\partial\phi}{\partial x}= 0\left\{\begin{array}{c l}                                                                              
\mbox{ $y<$0, $x=-x_1$ or $x=x_2$} \\
\end{array}
\right.
\end{equation}

\begin{equation}
\frac{\partial\phi}{\partial y}= 0\left\{\begin{array}{c l} 
\mbox{ $x<$$x_1$, $y$=0, $y$=$y_1$} \\
\end{array}
\right.
\end{equation}

\noindent
where -100$\lambda_F<<$0 and $|x_{lim}|$ the maximum propagation extent of the electron in x which is discussed further in the following section A. The solution for V=0.1mV, in addition to the electron paths taken through the system for V=0.01-1mV, can be seen in Fig. \ref{fig:dpath} and \ref{fig:dpath2}. Electron-electron interactions are not considered. We use this free electron method as the primary model for electron dynamics for all data shown here. 

\begin{figure}
\includegraphics[width=1.0\linewidth]{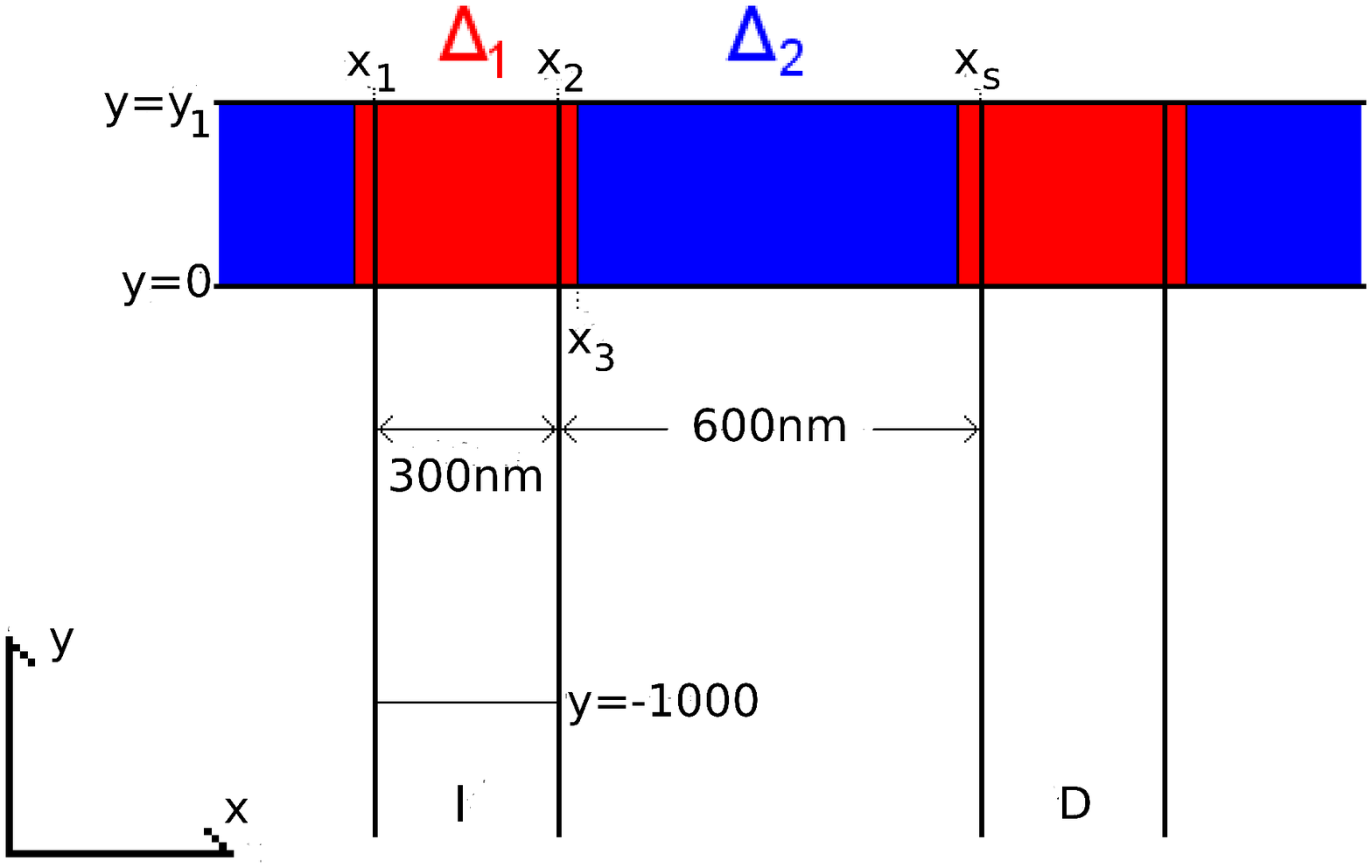}\\
\caption{Schematic of model configuration for a lateral spin valve geometry, with injector (I) and detector (D) electrodes indicated and inter-electrode separation L=600nm}
\label{fig:dm1}
\end{figure} 

\subsection{\label{sec:level2}Normal State}

In the normal state, an electron incident at $y$=0 is free to continue diffusive motion in the perpendicular electrode, with step size equal to the new mean free path $\lambda_S$ in the material. We define a position $|x_{lim}|$ whereby the electron is considered to have fully escaped the injector electrode region. For -$x_{lim}$ (to the left in Fig. \ref{fig:dm1}) this electron is considered to contribute to the local conductance; for +$x_{lim}$ the nonlocal. An electron exiting to the left is modeled to add +1 arbitrary conductance unit to the local total $\theta_l$ and exiting to the right adding +1 to the nonlocal total $\theta_r$. For N$_T$ single electrons passed through the system, the total conductance is defined as g$_l$ = $\sum_{N_T}\theta_l$ and similarly for g$_r$ = $\sum_{N_T}\theta_r$. 

For the case of inter-electrode separation L$>>$$\lambda_S$, in the normal state, the charge current flow in +$x$ towards the detector electrode should be zero should be zero as there is no path to ground, as a result a counter potential acting to drive electrons in the +x direction forms and experimentally it is this counter potential that is measured with a high impedance meter. This acts to drive electrons in -$x$, producing no net charge current flow as observed experimentally (Fig. \ref{fig:tmp1}). Experimentally, this potential is detected via measurement using a high impedance meter with no current flow into the detector electrode.  Without taking this`aspect into account, the model would produce an unphysically high value for g$_r$, especially as the applied injector bias V$\rightarrow$0. In this work we simulate the potential generated at the nonlocal electrode required to drive a second electron of equal and opposite velocity, obtained from the effect of a calculated counter potential (V$>$0) or from a distribution function (V=0), traveling back into the interface region (in the -$x$ direction). For summation over N$>$1 electrons, this results in no nonlocal conductance (g$_r$=0) in the normal state, matching the situation observed experimentally (Fig. \ref{fig:tmp1}).     

\begin{figure}
\includegraphics[width=1.0\linewidth]{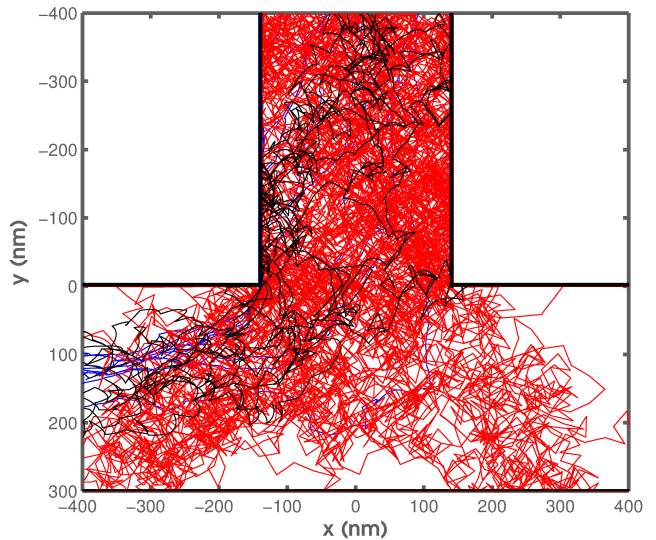}
\caption{(Color online) Electron paths in the normal state of the device for N=10 electrons and applied injector bias V=0.1mV (red),0.1mV (black) and 1mV (blue)}
\label{fig:dpath2}
\end{figure} 

Figure \ref{fig:dpath2} shows the paths taken by N$_{T}$=10 electrons for injector bias V=0.01-0.1mV, illustrating the operation of the model in the normal state. For low bias V=0.01mV, the electron may take a diffusive path in $\pm$x, indicated by the tracks towards +$x$ (right). For high bias, the majority of electrons move in -$x$, the paths confined closer to the inner left edge of the junction in the -$x$ direction as they approach $y$=0. For V=0, electron diffusion can occur with equal probability in either direction, zero net charge current in the +$x$ electrode imposed as a result of modeled injection of a counter electron moving in -$x$ (not shown) representing an opposing counter diffusion and ensuring g$_{r}$=0. However, for V=0 no net flow of electrons should occur to the injection region (i.e. N$_{T}$=0 at V=0), which would limit the validity of the model to finite bias in the normal state. This was accounted for in the model by fixing a finite maximum number of iterative time steps to $\leq$2x10$^{5}$, such that the electron never reaches x$_{lim}$ at V=0 in the normal state, giving g$_{r}$=g$_{l}\approx$0. 

\subsection{\label{sec:level2}Superconducting State}

\begin{table}
\centering
\begin{tabular}{ | l | l | l|}
\hline
Process & $\theta_l$, Local & $\theta_r$, Nonlocal\\
\hline
\hline
AR& +2 & 0 \\
CAR& +1 & +1\\
EC & +1 & -1\\
Normal reflection &-1&0\\
Quasiparticle(CI) &+1&$\pm$A$_q$e$^{-x_s/\Lambda_Q}$\\
\hline
\end{tabular}
\caption{Single-electron contributions from each process to local and nonlocal conductance with zero spin polarization P$_m$. At the nonlocal electrode CAR is defined to make a positive contribution (exhibited as a negative nonlocal resistance) and EC a negative one}
\label{tab:nummod1}
\end{table}

We next consider the superconducting state by introducing two spatially variant values of the superconducting gap $\Delta_1(x,y)$ and $\Delta_2(x,y)$ representing a suppressed and bulk energy gap, such that $\Delta_2(x,y)$ $>$ $\Delta_1(x,y)$. This follows the approach of Strijkers et al. \cite{2001PhRvB..63j4510S} considering a proximitized layer of lower-$\Delta$ at the normal-superconductor interface. For the model, $\Delta_1$=$\Delta_2$=0 for $y$$<$0 in the normal state electrodes. We restrict the spatial extent of the suppressed $\Delta_1$ to around the region of overlap between the normal electrodes and superconducting electrode for 0$<$$y$$<$$y_1$ and $|x_3|$ with $\Delta_2$ covering the remainder of the superconducting region, such that an electron may progress as a quasiparticle excitation in the $\Delta_1$ region for $\Delta_1$$<$E$<$$\Delta_2$ but not into the bulk $\Delta_2$ region. This is illustrated in Fig. \ref{fig:dm1}. Each electron is defined as possessing an energy E, taken at random from a Fermi distribution f(E-V) = 1/(e$^{(E-V)/k_bT}$+1) at each scattering step, where f$_p$ is a pseudo-random value (0..1), V the simulated potential across the injector and T the simulated temperature.

For an electron diffusively reaching the region of nonzero $\Delta$ and E$<$$\Delta_1$, only transmission by Andreev reflection is possible. The energy and potential dependent probability for AR P$_{AR}$ is calculated from expressions given by from the model of BTK \cite{1982PhRvB..25.4515B} and the modification by Strijkers. We consider that a fraction of this Andreev reflection may occur as CAR such that P$_{CAR}$=cP$_{AR}$ and that a fraction of the current which is not transmitted by AR could be by EC such that P$_{EC}$=c$_t$(1-P$_{AR}$+P$_{CAR}$) defining a probability conservation relation P$_{AR}$+P$_{CAR}$+P$_{NR}$+P$_{EC}$=1 including the probability of normal reflection P$_{NR}$. We assume a similar energy dependence for P$_{AR}$ and P$_{AR}$ and that the factors c and c$_t$ defining the CAR and EC fractions are invariant with electron energy or applied simulated potential V. For each modeled electron, the process undertaken is selected using a pseudo-random value p$_3$=(0..1), the range to 1 subdivided according to the ratios of probabilities for each process. An electron undertaking any of the processes converting to supercurrent is said to have exited the system, in a manner similar to that of diffusive exit past $|x_{lim}|$ and it's further diffusive motion is not considered - the contributions $\theta_{r/l}$ for each process are given in Table \ref{tab:nummod1}. 

For $\Delta_1$$<$E$<$$\Delta_2$ the electron may still Andreev reflect, cotunnel or may now enter the superconducting electrode as a quasiparticle. We consider the quasiparticle state to be electron-like and modeled by random walk diffusive motion - but limited to the $\Delta_1$ region within $|x_3|$, with quasiparticle transport into the bulk $\Delta_2$ region outside prohibited. For the calculation of the electron dynamics, we assume the electrostatic force on the quasielectron $\textbf{F}(x,y)$=0 within the superconducting regions, permitting motion with equal likelihood in all directions. Such a method for quasiparticle diffusion has been previously used in the literature \cite{warb}. 

For E$>$$\Delta_2$ the restriction to $|x_3|$ is removed, permitting quasiparticle diffusion throughout the superconducting electrode (whilst prohibiting any Andreev reflection). A quasiparticle reaching $|x_{lim}|$ contributes to both the local and nonlocal conductance according to Table \ref{tab:nummod1} with an exponential decay depending on the separation between the point of exit and the nonlocal electrode $x_s$, and decay scale $\Lambda_Q$ the charge imbalance length. This represents a phenomenological replication of the nonlocal potential produced by charge imbalance, of either negative or positive contribution to the nonlocal effect. The same dependence on inter-electrode separation is introduced for EC and CAR by reducing their fraction of the AR and non-AR probability at the injector according to an exponential form c=c$_0$e$^{-x_s/\xi_0}$ and c$_t$=c$_{t0}$e$^{-x_s/\xi_0}$ on a decay scale equal to the BCS coherence length $\xi_0$, c$_0$ and c$_t$ are defined at zero separation (i.e. - at the injector). Since E is taken from a Fermi distribution, even for low V all processes are possible at finite T. The spatial variation in $\Delta$ permits the simulation of regions of energy gap lower than the bulk value for the superconducting electrode. 

\subsection{\label{sec:level2}Spin Dependence}

Finally, we consider the case for a ferromagnetic injector and detector electrode by considering a total pool of available electrons of each spin type: N$_{1\uparrow}$, N$_{1\downarrow}$ at the injector and N$_{2\uparrow}$, N$_{2\downarrow}$, such that the total number of simulated injected electrons N$_{T}$=N$_{1\uparrow}$+N$_{1\downarrow}$ and defining the polarization for the spins in the model P$_m$ at the injector electrode:

\begin{equation}
P_m=\frac{(N_{1\uparrow}-N_{1\downarrow})}{(N_{1\uparrow}+N_{1\downarrow})}
\end{equation}

\noindent
P$_m$ is assumed to be uniform for both injector and detector and represents a means of replicating the bulk spin polarization P in the ferromagnetic electrode. A simulated electron is selected at random from one of the spin type pools (N$_{1(\uparrow/\downarrow)}$). For P$_m$=0 this has no effect on the conductance totals or probabilities for any effect. For finite P$_m$, such that N$_{1\uparrow}\ne$N$_{1\downarrow}$ and N$_{2\uparrow}\ne$N$_{2\downarrow}$, one pool will be exhausted before the other, prohibiting the spin dependent processes (such as AR, CAR) from occurring. For example, in the half metallic case where N$_{1\uparrow}$=N$_{2\uparrow}$=0, CAR would be prohibited due to the absence of the pairing spin-$\uparrow$ required to form a Cooper pair. This method also permits simulation of magnetization dependence by introducing an imbalance in the spin dependent pools such that N$_{1\uparrow}>$N$_{1\downarrow}$ and N$_{2\uparrow}<$N$_{2\downarrow}$ in the antiparallel configuration and the reverse for parallel magnetization. The model neglects spin accumulation and the non-local spin valve effect by choice, although such an effect could in principle be included. 

\subsection{\label{sec:level2}Model Examples}

\begin{figure}
\includegraphics[width=1.0\linewidth]{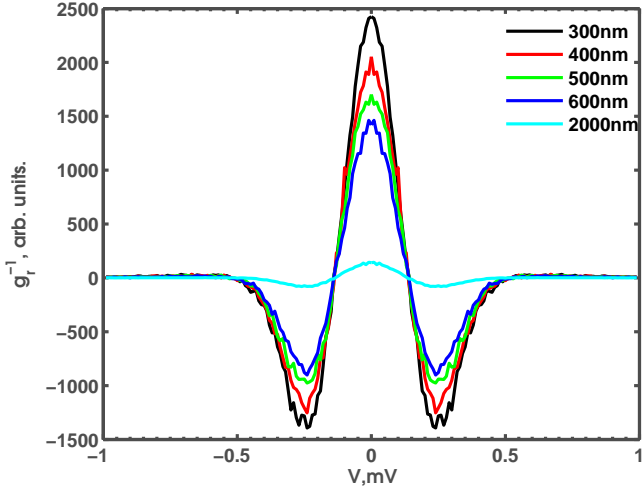}
\caption{Example of model output I:  nonlocal differential resistance g$_r^{-1}$ as a function of V for inter-electrode separation L=300-900nm.}
\label{fig:A}
\end{figure}

\begin{figure}
\includegraphics[width=1.0\linewidth]{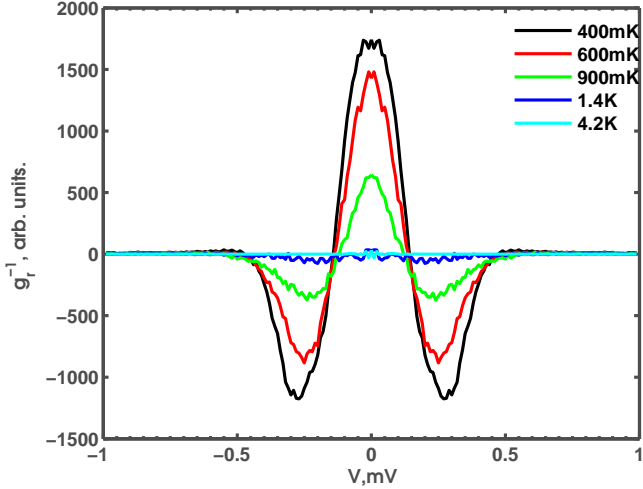}
\caption{Example of model output II: differential resistance g$_r^{-1}$ vs. V at a range of temperatures T=400mK-4.2K, with A$_q$=0 and parameters as in the main text}
\label{fig:B}
\end{figure}

\begin{figure}
\includegraphics[width=1.0\linewidth]{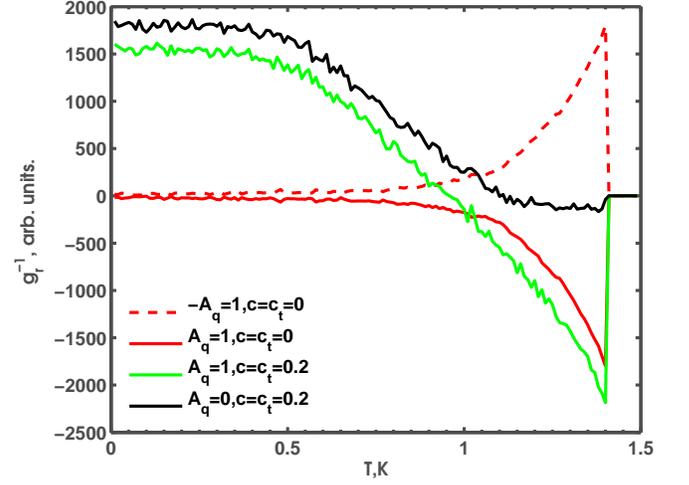}
\caption{Example of model output III: T-dependence for absence/presence of charge imbalance A$_q$=0 or $\pm$1 and absence/presence of EC/CAR (c,c$_t$ values) at V=0. The result for negative CI contribution (-A$_q$, solid) is shown with the mirrored result for positive contribution (+A$_q$, dashed)}
\label{fig:C}
\end{figure}

Figures \ref{fig:A}-\ref{fig:C} show the effects on the modeled nonlocal differential resistance g$_r^{-1}$ for an arbitrary selection of realistic parameters ($\xi_0$=600nm, $\Lambda_Q$=1$\mu$m, $\Delta_1$=0.2meV, $\Delta_2$=0.4meV, T=600mK, T$_c$=1.4K) as a function of inter electrode separation and of temperature for N$_{T}$=10000 electrons, both following qualitative trends observed by others \cite{2010Nanot..21A4002K}$^{,}$\cite{2010PhRvB..81b4515B}. Figure \ref{fig:B} can be compared directly to Figure \ref{fig:tmp1} from experiment, the same qualitative trend of reduction in peak and negative minima observed from the model.

Figure \ref{fig:C} shows the temperature dependence at V=0 ($I_{dc}$=0) for 3 configurations - no CI (A$_q$=0) and finite CAR/EC c=c$_t$=0.2 and with CI switched on: A$_q$=1 with absence/presence of CAR/EC. We consider the case $\Delta_1$=$\Delta_2$=0.2meV with a single T$_c$ of 1.4K. We model CI as a negative contribution to the nonlocal resistance g$_r^{-1}$ to assist replication of the effects observed experimentally. In the case of CI-only, no finite effect is seen at low-T. This did not match the result for the real Al device in Fig. \ref{fig:D} which qualitatively follows the model form with finite-CAR/EC contribution as T$\rightarrow$0 and CI divergence as T$\rightarrow$T$_c$. Divergent behavior is replicated as T$\rightarrow$T$_c$ in the cases where A$_q>0$ and a finite CI contribution is considered. Assumption of a negative CI contribution (A$_q$=1) combined with a finite positive low temperature CAR/EC contribution produces a negative divergence towards T$_c$ but of a form not reflected in experimental data. We also note that assumption of a positive contribution to the nonlocal resistance (negative A$_q$) with increasing T, the behavior for A$_q$=-1, c=c$_t$=0 (CI only, dashed line in Figure \ref{fig:C}) replicates the observed behavior seen by others \cite{2010Nanot..21A4002K}$^{,}$\cite{2010PhRvB..81r4524H}$^{,}$\cite{2004PhRvL..93s7003B} as an increase to a finite peak from low-T zero g$_r^{-1}$ with zero effect above T$_c$. 

\begin{figure}
\includegraphics[width=1.0\linewidth]{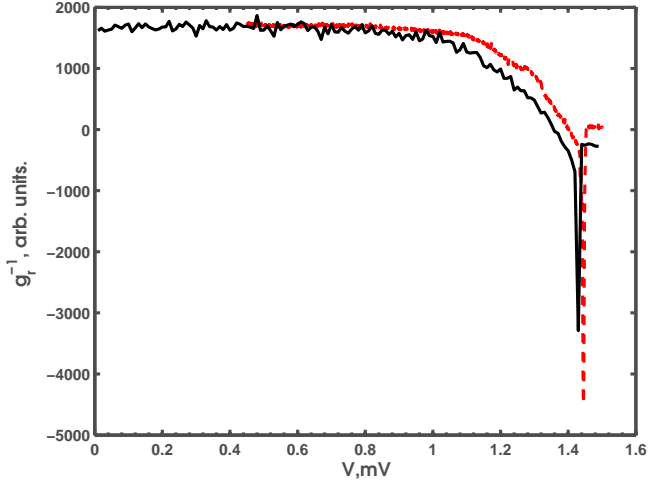}
\caption{Example of model output IV: (solid) the case where the region -$x_1$$<$$x$$<$$x_2$, $y$$>$0 enters the normal state at T=T$_{c}^{B}$-0.04K with T$_{c}^{B}$=1.4K the bulk critical temperature of the $\Delta_2$ region which remains superconducting. In this case, transport may occur by direct quasiparticle transmission (or minimal AR) at the $\Delta_2$ interface in $\pm x$. The result is a sharp, negative minima qualitatively matching the effect observed experimentally (dashed, scaled from Fig. \ref{fig:D})}
\label{fig:special}
\end{figure}

Figure \ref{fig:special} shows the modeled case whereby the T$_{c}$=T$_{c}^{A}$=1.4K of the $\Delta_1$(T=0)=0.213meV region is set 0.04K lower than the bulk T$_{c}^{B}$=1.44K of the $\Delta_2$(0)=0.22meV region, such that for T$_{c}^{A}$$<$T$<$T$_{c}^{B}$ a fully normal state region is formed at the overlap between the ferromagnetic and superconducting electrodes. This would be anticipated experimentally, especially in the case of Ohmic contacts such as utilized in devices studied in this work, due to the suppression of superconductivity at the F/S interface. Within this temperature range, model nonlocal processes such as CAR and EC are limited due to the low gap energy in the high-T regime. At low bias within the interface region, diffusive electron motion permits incidence with the $\Delta_2$ interface at either -$x_3$ or +$x_3$. An electron arriving at the +$x_3$ interface may enter the superconductor primarily as a quasiparticle excitation close to T$_c$ (due to the low value of $\Delta_2$) and with energy E$>$$\Delta_2$ - we define this electron as producing a negative nonlocal contribution to g$_{r}^{-1}$ of magnitude $|$A$_q|$e$^{-x_s/\Lambda_Q}$. In contrast, for the fully normal state $\Delta_1$=$\Delta_2$=0 CI cannot occur to produce a nonequilibrium population of electrons and a detectable potential difference, as discussed in section A. A combination of the reduction in $x_s$, a higher number of electrons reaching +$x_3$ and the proximity to bulk T$_{c}^{B}$ producing a longer $\Lambda_Q$ and larger CI effect results in a negative effect that can be far greater than that produced by nonlocal CI. This provides a potential explanation for the divergent features observed in experiment as plotted in Fig. \ref{fig:D}, with which excellent qualitative agreement is obtained. 

\subsection{\label{sec:level2}Match to Experiment}

\begin{figure}
\includegraphics[width=1.0\linewidth]{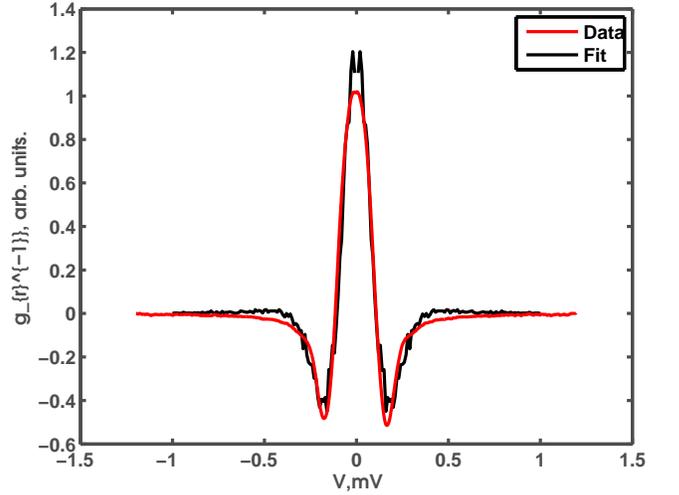}\\
\caption{Nonlocal differential resistance as a function of applied injector DC current $I_{dc}$ for modeled and experimental data. Device inter-electrode separation was L=600nm and experimental measurement temperature 600mK. }
\label{fig:comp}
\end{figure}

In order to demonstrate the capability of the model, the transport behavior of the real device was compared to the model result with input parameters realistic for the fabricated devices. The result of this is shown in Figure \ref{fig:comp} giving the model result overlaid onto real nonlocal differential resistance data from experiment. Model parameters were selected to best represent the device or from measurement of device properties. Parameters $x_1$=-150nm, $x_2$=+150nm (giving width=300nm) were fixed based on the physical dimensions of the device. The point of diffusive exit from the system $|x_{lim}|$=450nm, taken as the half-way point to the detector electrode with inter-electrode separation L=600nm, based on the assumption arrived at through simulation that the majority of electrons crossing past $x$=300nm would follow a path reaching the second electrode at $x$=750nm. $|x_{lim}|$ was also minimized to reduce computation time, but to be sufficiently high to inhibit premature exit of the electron from the system. 

Values of $\lambda_s$=15nm, $\lambda_F$=5nm and $\Lambda_Q$=1$\mu$m were used, based on physically reasonable parameters from previous work for nonlocal spin value devices and on nonlocal charge imbalance. $\xi_0$=500nm was taken, assuming a relatively clean superconducting Al electrode, but with a degree of suppression of the coherence length from the BCS value and sufficiently long to produce non-negligible superconducting nonlocal effects over the inter-electrode separation used. Although a value was set for $\Lambda_Q$, based on the temperature independence of the nonlocal effect at the 600mK experimental measurement temperature (Fig. \ref{fig:D}) the assumption was made of complete suppression of nonlocal charge imbalance. We explicitly introduce the absence of CI by setting A$_q$=0, in order to demonstrate that the effects observed can originate solely from non-CI processes. The assumption that CI can be completely eliminated in this manner was based on the hypothesis from experimental data that the origin of the nonlocal effect at T$<<$T$_c$ was from processes operating over a decay scale $\xi_0$ (i.e. - CAR/EC) rather than CI. This assumption would not be valid for T$\rightarrow$T$_c$, where additional CI effects would arise from the divergence of $\Lambda_Q$ and cannot replicate the temperature dependent effect observed experimentally. 

Modeled T$_c$ was set to 1.4K, close to that observed in experimental data. Using this value of T$_c$, we calculated physically reasonable values for $\Delta_1$=0.112meV, $\Delta_2$(0)=0.301meV derived from standard BCS theory and fits to local (single junction) conductance data. We assume the simplest case with uniform values of $\Delta_1$ and $\Delta_2$ for injector and detector junctions to the superconductor. The requirement for the two values of $\Delta$ was based on the single junction experimental data, whereby a region of lower energy gap was required to correctly fit the conductance behavior - this region likely being that immediately adjacent to the magnetic electrodes. On this basis we limit the $\Delta_1$ region extent $|x_3|$ to within the overlap region of the injector/detector electrodes -150$<$$x$$<$150nm . We neglect any proximity effect into the injector, confining the superconducting region of nonzero $\Delta$ strictly to within the perpendicular lateral electrode for 300$<y<$0nm. 

The values of c and c$_t$ represented free parameters which were selected to give the best fit to the experimental data, each parameter effectively controlling negative minima depth and zero bias peak height of nonlocal resistance g$_r^{-1}$ respectively. We make the assumption that both c and c$_t$ are invariant with other parameters such as injector bias. Values of c=0.28 and c$_t$=0.15 were found to give a best fit, implying the presence of both effects but an imbalance in a 2:1 ratio between them such as not to cancel each other, with each producing a different effect at varying simulated bias. 

In order to translate the arbitrary conductance units used in the model to real nonlocal resistance (in Ohms) the conversion g$_r$=k$\delta$R$_{nl}$ was used with k a scaling factor of approximately 1000. A total of N$_T$=10000 electrons were simulated for each step in simulated injector V. A good fit was found to the experimental conductance data. As can be seen in the figure, the only discrepancies arose in matching the height of the peak, potentially due to a slightly high values of c or c$_t$, and a discontinuity around 0.23-0.3mV due to the 2-$\Delta$ approach taken, rather than a continuous range of $\Delta$ arising from the proximity effect. The fit was obtained with nonlocal charge imbalance explicitly removed (A$_q$=0), in order to demonstrate the absence of such effects at low temperature.  

\section{\label{sec:level1}Conclusion}

In conclusion, we have fabricated Al/Py devices suitable for measurement of nonlocal superconducting effects. We observe behavior characteristic of EC or CAR in the low temperature regime and nonlocal charge imbalance, arising as divergence in nonlocal differential resistance, close to T$_c$. In order to further investigate there results we have developed a simple numerical model of nonlocal processes in a lateral spin valve geometry device that is capable of qualitatively replicating the nonlocal effect trends with T and bias in experimental work here and in the literature, associated with the superconducting processes. The model is capable of quantitatively matching the nonlocal resistance from a real Al/Py device, using realistic parameters for the model variables. The single electron approach offers potential extension for shot noise calculation, of interest for recent cross correlation and entanglement studies. Although possibly inferior to a true analytical model, the numerical approach encompasses the key aspects of the physics of such a device and gives insight and the ability to replicate the typical local properties of a device and provides a simple way to derive important parameters such as $\xi_0$ and $\Lambda_Q$.

\section{\label{sec:level1}Acknowledgements}

This work was supported by a UK Engineering and Physical Sciences Research Council Advanced Research Fellowship EP/D072158/1.

\bibliography{mega2}

\begin{thebibliography}{10}

\bibitem{2004PhRvL..93s7003B}
D.~{Beckmann}, H.~B. {Weber}, and H.~{v.~L{\"o}hneysen},
\newblock Physical Review Letters {\bf 93}, 197003 (2004).

\bibitem{2005PhRvL..95b7002R}
S.~{Russo}, M.~{Kroug}, T.~M. {Klapwijk}, and A.~F. {Morpurgo},
\newblock Physical Review Letters {\bf 95}, 027002 (2005).

\bibitem{2007ApPhA..89..603B}
D.~{Beckmann} and H.~{v.~L{\"o}hneysen},
\newblock Applied Physics A: Materials Science \& Processing {\bf 89}, 603
  (2007).

\bibitem{2010NatPh...6..494W}
J.~{Wei} and V.~{Chandrasekhar},
\newblock Nature Physics {\bf 6}, 494 (2010).

\bibitem{2009NatPh...5..393C}
P.~{Cadden-Zimansky}, J.~{Wei}, and V.~{Chandrasekhar},
\newblock Nature Physics {\bf 5}, 393 (2009).

\bibitem{2010PhRvB..81r4524H}
F.~{H{\"u}bler}, J.~C. {Lemyre}, D.~{Beckmann}, and H.~{v.~L{\"o}hneysen},
\newblock Physical Review B {\bf 81}, 184524 (2010).

\bibitem{2000ApPhL..76..487D}
G.~{Deutscher} and D.~{Feinberg},
\newblock Applied Physics Letters {\bf 76}, 487 (2000).

\bibitem{1982PhRvB..25.4515B}
G.~E. {Blonder}, M.~{Tinkham}, and T.~M. {Klapwijk},
\newblock Physical Review B {\bf 25}, 4515 (1982).

\bibitem{PhysRevB.68.174504}
T.~Yamashita, S.~Takahashi, and S.~Maekawa,
\newblock Phys. Rev. B {\bf 68}, 174504 (2003).

\bibitem{2006PhRvB..74u4512B}
A.~{Brinkman} and A.~A. {Golubov},
\newblock Physical Review B {\bf 74}, 214512 (2006).

\bibitem{2009PhRvL.103f7006G}
D.~S. {Golubev}, M.~S. {Kalenkov}, and A.~D. {Zaikin},
\newblock Physical Review Letters {\bf 103}, 067006 (2009).

\bibitem{jr.:4589}
J.~R.~J.~Soulen et~al.,
\newblock Journal of Applied Physics {\bf 85}, 4589 (1999).

\bibitem{2001PhRvB..63j4510S}
G.~J. {Strijkers}, Y.~{Ji}, F.~Y. {Yang}, C.~L. {Chien}, and J.~M. {Byers},
\newblock Physical Review B {\bf 63}, 104510 (2001).

\bibitem{PhysRevLett.95.027002}
S.~Russo, M.~Kroug, T.~M. Klapwijk, and A.~F. Morpurgo,
\newblock Phys. Rev. Lett. {\bf 95}, 027002 (2005).

\bibitem{2010PhRvB..81b4515B}
J.~{Brauer}, F.~{H{\"u}bler}, M.~{Smetanin}, D.~{Beckmann}, and
  H.~{v.~L{\"o}hneysen},
\newblock Physical Review B {\bf 81}, 024515 (2010).

\bibitem{2006PhRvL..97w7003C}
P.~{Cadden-Zimansky} and V.~{Chandrasekhar},
\newblock Physical Review Letters {\bf 97}, 237003 (2006).

\bibitem{2007NJPh....9..116C}
P.~{Cadden-Zimansky}, Z.~{Jiang}, and V.~{Chandrasekhar},
\newblock New Journal of Physics {\bf 9}, 116 (2007).

\bibitem{2010Nanot..21A4002K}
A.~{Kleine} et~al.,
\newblock Nanotechnology {\bf 21}, A264002 (2010).

\bibitem{warb}
P.~Warburton and M.~Blamire,
\newblock Applied Superconductivity, IEEE Transactions on {\bf 5}, 3022
  (1995).

\end{thebibliography}

\end{document}